\documentclass[aps,prl,twocolumn,preprintnumbers,superscriptaddress,amsmath]{revtex4}

\usepackage{amssymb}
\usepackage{txfonts}
\usepackage{amssymb}
\usepackage{graphicx}
\usepackage{subfigure}
\usepackage{dcolumn}
\usepackage{bm}
\usepackage{color}
\usepackage{upgreek}

\begin{document}

\title{Transverse mode-encoded quantum gate on a silicon photonic chip}

\author{Lan-Tian Feng\footnote[1]{These authors contributed equally to this work.}}
\affiliation
{Key Laboratory of Quantum Information, CAS, University of Science and Technology of China, Hefei 230026, China.}
\affiliation{Synergetic Innovation Center of Quantum Information $\&$ Quantum Physics, University of Science and Technology of China, Hefei 230026, China.}
\author{Ming Zhang\footnote[1]}
\affiliation{State Key Laboratory for Modern Optical Instrumentation, Centre for Optical and Electromagnetic Research, Zhejiang Provincial Key Laboratory for Sensing Technologies, Zhejiang University, Zijingang Campus, Hangzhou 310058, China.}
\affiliation{Ningbo Research Institute, Zhejiang University, Ningbo 315100, China.}
\author{Xiao Xiong}
\author{Di Liu}
\author{Yu-Jie Cheng}
\author{Fang-Ming Jing}
\author{Xiao-Zhuo Qi}
\author{Yang Chen}
\author{De-Yong He}
\author{Guo-Ping Guo}
\author{Guang-Can Guo}
\affiliation
{Key Laboratory of Quantum Information, CAS, University of Science and Technology of China, Hefei 230026, China.}
\affiliation{Synergetic Innovation Center of Quantum Information $\&$ Quantum Physics, University of Science and Technology of China, Hefei 230026, China.}
\author{Dao-Xin~Dai\footnote[2]{dxdai@zju.edu.cn}}
\affiliation{State Key Laboratory for Modern Optical Instrumentation, Centre for Optical and Electromagnetic Research, Zhejiang Provincial Key Laboratory for Sensing Technologies, Zhejiang University, Zijingang Campus, Hangzhou 310058, China.}
\affiliation{Ningbo Research Institute, Zhejiang University, Ningbo 315100, China.}
\author{Xi-Feng Ren\footnote[3]{renxf@ustc.edu.cn}}
\affiliation
{Key Laboratory of Quantum Information, CAS, University of Science and Technology of China, Hefei 230026, China.}
\affiliation{Synergetic Innovation Center of Quantum Information $\&$ Quantum Physics, University of Science and Technology of China, Hefei 230026, China.}

\begin{abstract}
As an important degree of freedom (DoF) in integrated photonic circuits, the orthogonal transverse mode provides a promising and flexible way to increasing communication capability, for both classical and quantum information processing. To construct large-scale on-chip multimode multi-DoF quantum systems, a transverse mode-encoded controlled-NOT (CNOT) gate is necessary. Here, through design and integrate transverse mode-dependent directional coupler and attenuators on a silicon photonic chip, we demonstrate the first multimode implementation of a two-qubit quantum gate. With the aid of state preparation and analysis parts, we show the ability of the gate to entangle two separated transverse mode qubits with an average fidelity of $0.89\pm0.02$ and the achievement of 10 standard deviations of violations in the quantum nonlocality verification. In addition, a fidelity of $0.82\pm0.01$ was obtained from quantum process tomography used to completely characterize the CNOT gate. Our work paves the way for universal transverse mode-encoded quantum operations and large-scale multimode multi-DoF quantum systems.
\end{abstract}
\maketitle

In integrated optics, a significant enhancement of information processing and communication capability is an ultimate goal due to the exponential growth in optical interconnection and communication requirements for both classical \cite{Dai2013} and quantum \cite{Wang2018} information applications. To achieve large-scale photonic quantum systems, multiphoton, multiple degrees of freedom (DoFs) and high-dimensional encoding is becoming attractive and essential \cite{Wang2019}. In recent years, a lot of technical and experimental efforts to achieve universal quantum operations have been done with different DoFs, such as single qubit operation and two-qubit controlled-NOT (CNOT) gate with path or polarization encoding \cite{O'Brien2009,O'Brien2003,Kiesel2005,Politi2008,Crespi2011,Shadbolt2012}, as well as two-qubit quantum gate for time-bin qubits \cite{Humphreys2013,Lo2020} and frequency-bin qubits \cite{Lu2019}. In addition, multi-DoF hybrid encoding has also shown impressive applications such as super-dense coding \cite{Barreiro2008} and multi-dimensional teleportation \cite{Wang2015}.

More recently, the orthogonal transverse modes supported by one multimode waveguide are valued and developed rapidly for their high-dimensional scalability, compactness and coherent conversion with other DoFs \cite{Li2019,Luo2014,Dai2018,Feng2016}. As a new degree of freedom, transverse mode provides the flexibility and scalability for a wide scope of novel applications, such as dense optical interconnection \cite{Dai2013,Luo2014,Dai2018}, quantum information science \cite{Feng2016,Mohanty2017,Feng2019,Paesani2020}, nonlinear photonics \cite{Wang2017,Kittlaus2017}, and so on. Meanwhile, a variety of multimode devices such as mode (de)multiplexer \cite{Ding2013,Wang20142}, grating coupler \cite{Lai2018}, switch \cite{Stern2015,Jia2017,Xiong2017}, filter \cite{Huang2016}, sharp bending and mode-independent crossing \cite{Liu2019,Li2018}, and many other building blocks \cite{Dai2021} have been developed intensively. However, as a potential on-chip DoF, the full manipulation of transverse mode qubits, especially a two-qubit CNOT gate, is still not studied extensively and challenging.

\begin{figure*}[t]
\centering
\includegraphics[width=14cm]{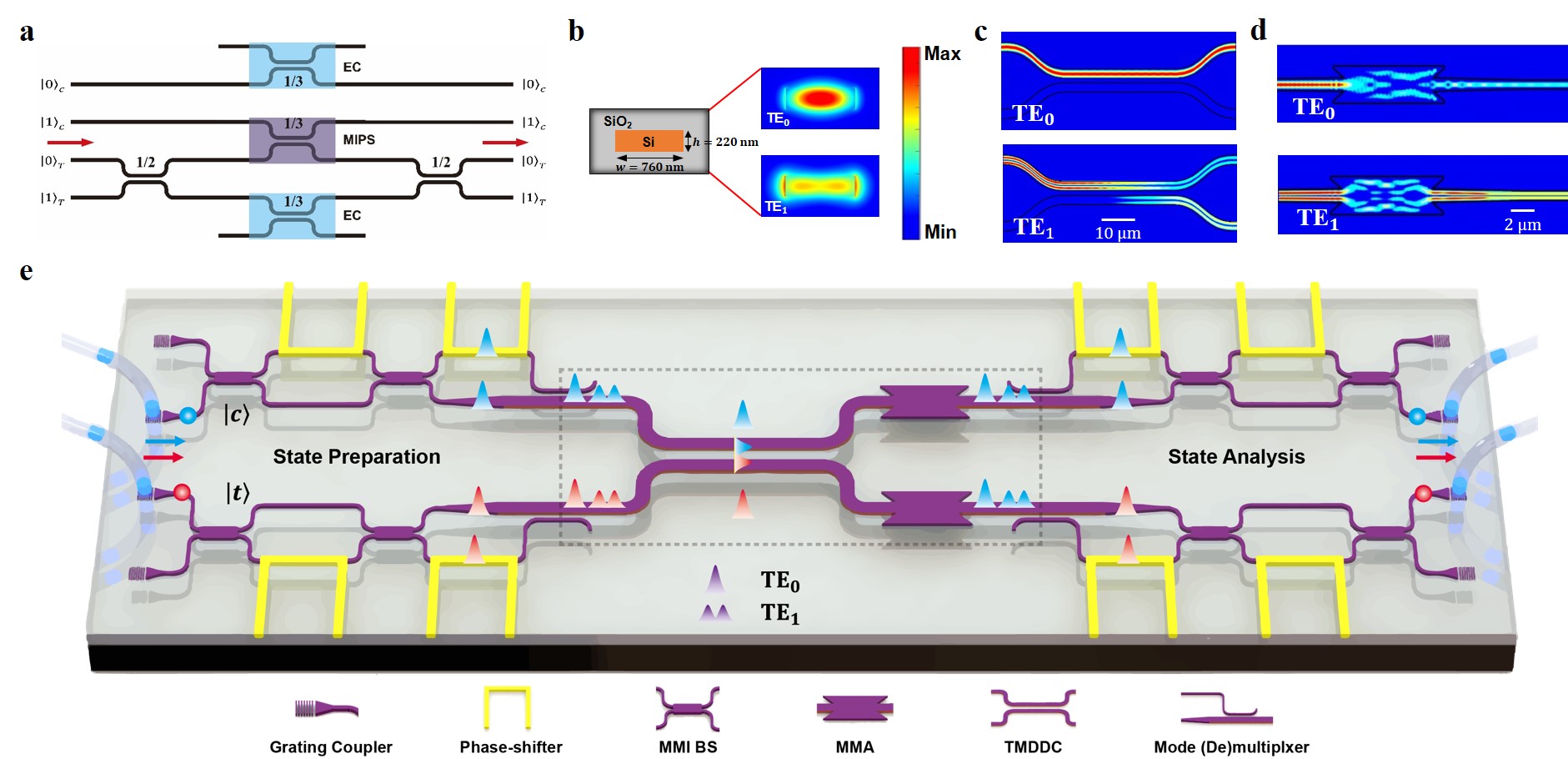}
\caption {Transverse mode-encoded quantum CNOT gate. (a) Common circuit implementation of the path-encoded quantum CNOT gate. MIPS: measurement-induced controlled phase; EC: energy compensator. (b) The transverse mode in a multimode waveguide can be directly used for quantum information encoding in a compact manner. (c) Simulated light propagation in the designed TMDDC. As the two identical multimode waveguides gradually merge, the TE$_0$ mode remains in the top waveguide without any coupling, while the TE$_1$ mode is cross-coupled with a ratio of 2/3 from the upper waveguide to the lower one. (d) Simulated light propagation in the designed MMA. (e) Schematic of the entire silicon photonic integrated circuit. Photon pairs generated in free space are coupled to the chip through a fiber array. After state evolution in the chip, the photon pairs are coupled to another fiber array and received by single-photon detectors. The dashed black frame marks out the multimode section, where both modes are supported and function together as a transverse mode encoded CNOT gate. The other parts of the circuits are single-mode and used for state preparation and analysis. MMI BS: multimode interference beam splitter; MMA: multimode attenuator; TMDDC: transverse mode dependent directional coupler.}
\label{fig1}
\end{figure*}

In this paper, we report the on-chip generation, manipulation, and measurement of transverse mode-encoded quantum entangled states based on a compact CNOT gate. The CNOT gate relies on two newly developed multimode devices that can selectively manipulate different transverse modes, called unbalanced transverse mode-dependent directional coupler (TMDDC) and multimode attenuators (MMAs). This multimode waveguide CNOT gate is very compact and has a size of approximately 10$\times$160 $\upmu$m$^2$. We characterize it by performing quantum state tomography and quantum process tomography measurements. The gate realizes the entanglement of two separable qubits with an average state fidelity of $\bar{F}_s=0.89\pm0.02$ and a process fidelity of $F_p=0.82\pm0.01$. Because transverse mode-encoding is compatible with other encoding protocols and facilitates the construction of quantum photonic integrated circuits (QPICs) with higher dimensions and capacity, our work paves the way for universal transverse mode-encoded quantum operations and practical applications in compact and scalable photonic quantum processing with multiple DoFs.

A quantum processor requires a cascade of various unitary operations, which consist of a finite series of basic logic gates. It has been proven that a two-qubit CNOT gate, which flips the target qubit $\left|t\right\rangle$ depending on the state of the control qubit $\left|c\right\rangle$,  together with single-qubit logic gates, is sufficient for universal quantum operation \cite{Nielsen2000}. In QPICs, the universal quantum operation is always implemented using beam splitters and phase controllers \cite{O'Brien2009,Politi2008,Crespi2011,Shadbolt2012}. The common circuit implementation of the quantum CNOT gate is shown in Fig. 1a. It is mainly composed of three unbalanced directional couplers, one for quantum interference to induce the controlled-phase (CPhase) operation, and the other two for energy compensation. Two 50/50 beam splitters are used to transform the CPhase operation to the CNOT operation by applying the relation $\mathrm{CNOT}=(I\otimes H)\mathrm{CPhase}(I\otimes H)$, where $I$ and $H$ denote the identity and Hadamard gates. This transformation can also be easily achieved by selecting the computation basis of the target qubit, as demonstrated in Ref. \cite{Crespi2011}.


Among the various material systems for QPICs, silicon-on-insulator was chosen here because of its excellent optical properties and CMOS compatibility \cite{Bogaerts2005}. The designed multimode waveguide has a cross-section of $760\times 220~\textrm{nm}^2$ and supports the two lowest-order TE polarization transverse modes, that is, the TE$_0$ and TE$_1$ modes, as shown in Fig. 1b. The modes are orthogonal to each other and no cross-coupling between them, thus they can be directly used for quantum information encoding. We have newly designed two mode dependent devices, TMDDC and MMA, which can achieve mode selective coupling and loss, as shown in Fig. 1c and 1d. By cascading a TMDDC and two MMAs, we constructed one multimode two-qubit CNOT gate, Fig. 1e.

The TMMDC was based on a symmetric directional coupler, in which the incident modes (i.e., TE$_0$, TE$_1$) can be partially transferred to the adjacent waveguide by evanescent coupling. Here, to achieve a mode selection beam splitter, we chose a wide waveguide width $w_0 = 760~\textrm{nm}$ and a large gap $g = 400~\textrm{nm}$. In this situation, the evanescent field of TE$_0$ mode will be restricted in the waveguide with barely no power transfer, but the TE$_1$ mode still has enough coupling strength and a coupling length $L = 32.5~$\textmu m was used to realize a cross-coupling ratio of 2/3. Fig. 1(c) shows the simulated light propagation in the designed structure when the TE$_0$ and TE$_1$ modes were launched.
The MMA was designed to attenuate 2/3 power of the TE$_0$ mode and guarantee lossless propagation of TE$_1$ mode for energy compensation. In order to realize such an energy attenuator, a multimode interferometer (MMI) was introduced due to its compact size ( \textless 3$\times$10 \textmu m$^2$) and low insertion losses (IL \textless 0.5 dB). For the lossless propagation of TE$_1$, we set the MMI length determined by the interference-induced self-imaging of TE$_1$ mode. In this case, the TE$_0$ mode shows imperfect self-imaging and the output energy depends on the device width. Here, we chose a device width $w = 2.5~$ \textmu m and length $l = 7.5~$ \textmu m, which will maintain lossless propagation of TE$_1$ mode while attenuating 2/3 of the power of TE$_0$ mode. In order to radiate away the unwanted energy left in the attenuator which would cause unnecessary interference and negatively affect the output, we introduced a butterfly-shape MMI to reduce reflection of TE$_0$ mode by adding 4 acute angles to the MMI. Fig. 1(d) shows the simulated light propagation in the designed attenuator when the TE$_0$ and TE$_1$ modes were launched. For more information on these two structures, see the Supplementary Materials (Section I). The clever design of controlling transverse modes separately here enables us to further cascading them for realizing complex functions with mode encoding such as two-qubit CNOT gate.

In the experiment, a PPKTP crystal was used to generate frequency-degenerate photon pairs centered at wavelength of 1540 nm. More details are provided in the Supplementary Materials (Section II). The photon pairs were collected with a single-mode fiber array and then coupled into the TE$_0$ mode in the on-chip single-mode waveguides via grating couplers. To encode photons with different transverse modes, mode (de)multiplexers \cite{Dai2013} were introduced before (after) the photons have passed through the CNOT gate. After state evolution in the chip, the photons were coupled using another fiber array and detected with superconducting nanowire single-photon detectors for the coincidence measurement. To prepare and detect the different quantum states, path encoding Mach-Zehnder (MZ) interferometers were performed. Micro-heaters controlled by a homemade multi-channel direct-current regulated power supply were used to thermally tune the phases of the photons. All the heaters were individually characterized and displayed high quality. See the Supplementary Materials (Section IV) for further details.


To characterize the chip with different input qubits, we defined logical and physical qubits with the following states: $\left|0\right\rangle_c=\left|\mathrm{TE}_0\right\rangle$, $\left|1\right\rangle_c=\left|\mathrm{TE}_1\right\rangle$, $\left|0\right\rangle_t=(\left|\mathrm{TE}_0\right\rangle+\left|\mathrm{TE}_1\right\rangle)/\sqrt{2}$, $\left|1\right\rangle_t=(\left|\mathrm{TE}_0\right\rangle-\left|\mathrm{TE}_1\right\rangle)/\sqrt{2}$, where the subscripts $c$ and $t$ denote the control and target qubits. This configuration has a success probability of $P=1/9$, and it can be rendered deterministic by making use of ancillary resources, measurements, and feed-forward \cite{Knill2001}.


\begin{figure}[t]
\centering
\includegraphics[width=4cm]{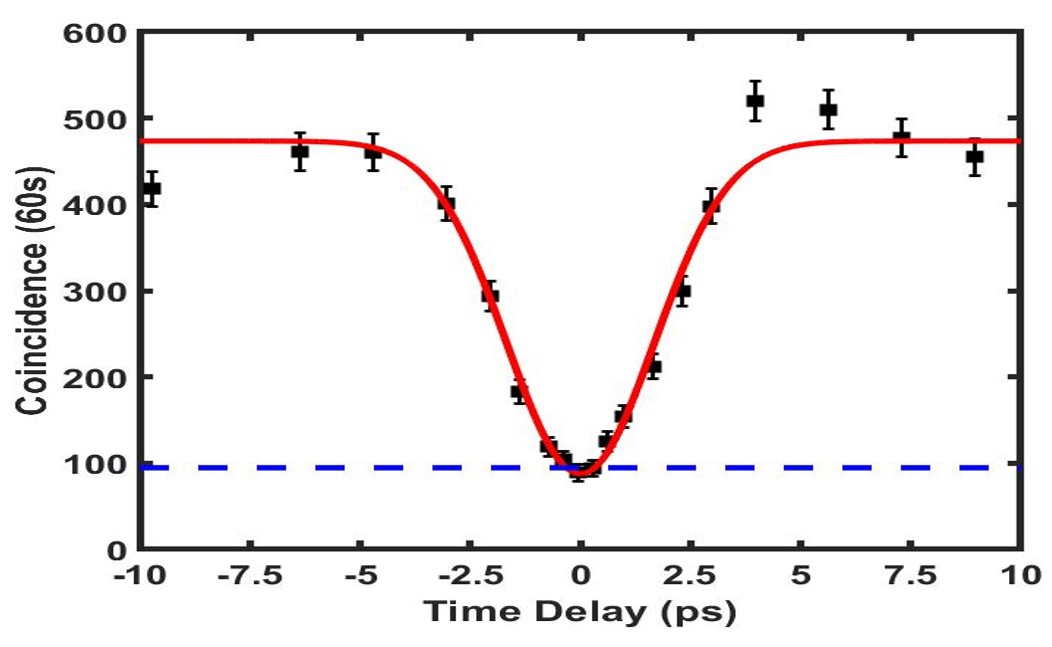}
\caption {Quantum interference in the TMDDC. Both the control and target photons are in the TE$_1$ mode and have an adjustable time delay between them. The dashed blue line denotes the lower limit for an ideal quantum interferometer with 1:2 split ratio. The error bars represent $\pm\sqrt{N}$ for the raw coincidence data (black rectangles) $N$. The dots denote the experimental data, and the curves the corresponding Gaussian fit.}
\label{fig2}
\end{figure}

Two-photon quantum interference is essential in the CNOT gate and the gate operation succeeds when two photons arrive at the directional coupler at the same time, thus the performance of the CNOT gate is directly related to the interference visibility. We first characterized the quantum interference in the multimode directional coupler. The input/output control and target photons were $\left|c\right\rangle=\left|\mathrm{TE}_1\right\rangle$ and $\left|t\right\rangle=\left|\mathrm{TE}_1\right\rangle$, respectively. By adjusting the time delay between the two photons, a quantum interference fringe was observed, as shown in Fig. 2. An interference visibility of $0.82\pm0.03$ was derived from Gaussian fitting before background subtraction (the raw data used, the same below) where the error was the fitting standard error. Here, the visibility is defined as $V=(C_{\rm{max}}-C_{\rm{min}})/C_{\rm{max}}$, where $C_{\rm{max}}$ and $C_{\rm{min}}$ are the maximum and minimum of fitted data, respectively. For the configuration in Fig. 1(a), the quantum interference should ideally give $C_{\rm{min}}=0.2C_{\rm{max}}$, and thus the ideal interference visibility is 0.80, as indicated by the dashed blue line in Fig. 2. The excellent agreement between the measured and ideal values for the quantum interference visibility proves that the TMDDC works well. In principle, the CNOT gate operation between the control and target qubits can be realized at zero time delay (dip position).

\begin{figure}[t]
\centering
\includegraphics[width=7cm]{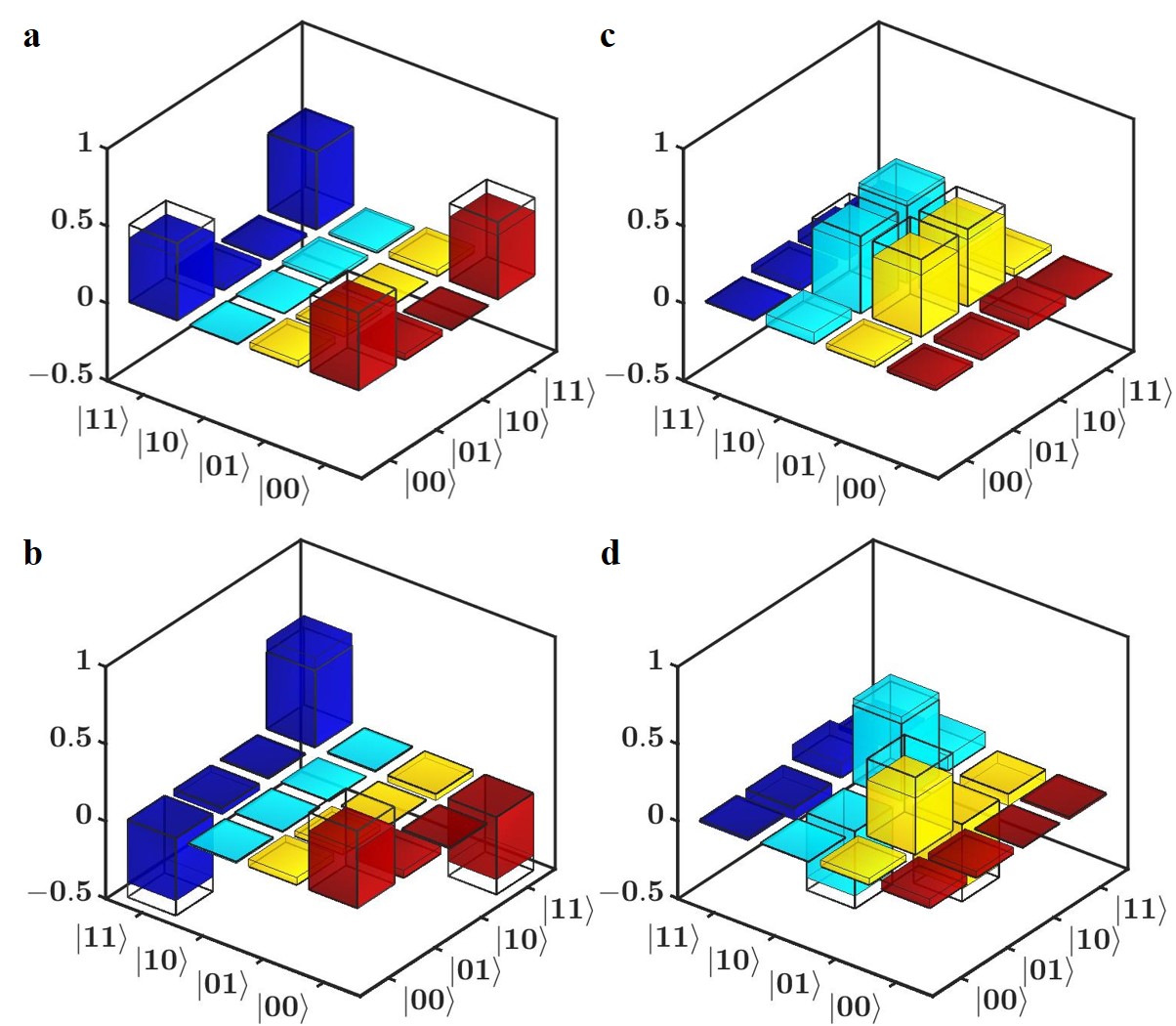}
\caption {Quantum state tomography of the generated Bell states. The real parts of the measured density matrices of the output states with the different input states of (a) $\left|+\right\rangle_c\left|0\right\rangle_t$, (b) $\left|-\right\rangle_c\left|0\right\rangle_t$, (c) $\left|+\right\rangle_c\left|1\right\rangle_t$, and (d) $\left|-\right\rangle_c\left|1\right\rangle_t$. The imaginary parts of the measured density matrices are negligible. The clear bars represent the ideal density matrices for the maximally entangled Bell states $|\Phi^{\pm}\rangle$ and $|\Psi^{\pm}\rangle$.}
\label{fig3}
\end{figure}


The basic function of a CNOT gate is entangling two separate qubits. This is one fundamental operation in quantum information processing and represents the most non-classical implication of quantum mechanics. Here, we use this operation to characterize the CNOT gate with an input of $\{\left|\pm\right\rangle_c\left|0\right\rangle_t, \left|\pm\right\rangle_c\left|1\right\rangle_t\}$ where $\left|+\right\rangle_c=(\left|0\right\rangle_c+\left|1\right\rangle_c)/\sqrt{2}$ and $\left|-\right\rangle_c=(\left|0\right\rangle_c-\left|1\right\rangle_c)/\sqrt{2}$. The output states were collected on-chip and reconstructed by quantum state tomography \cite{James2001}.

\begin{figure*}[t]
\centering
\includegraphics[width=12cm]{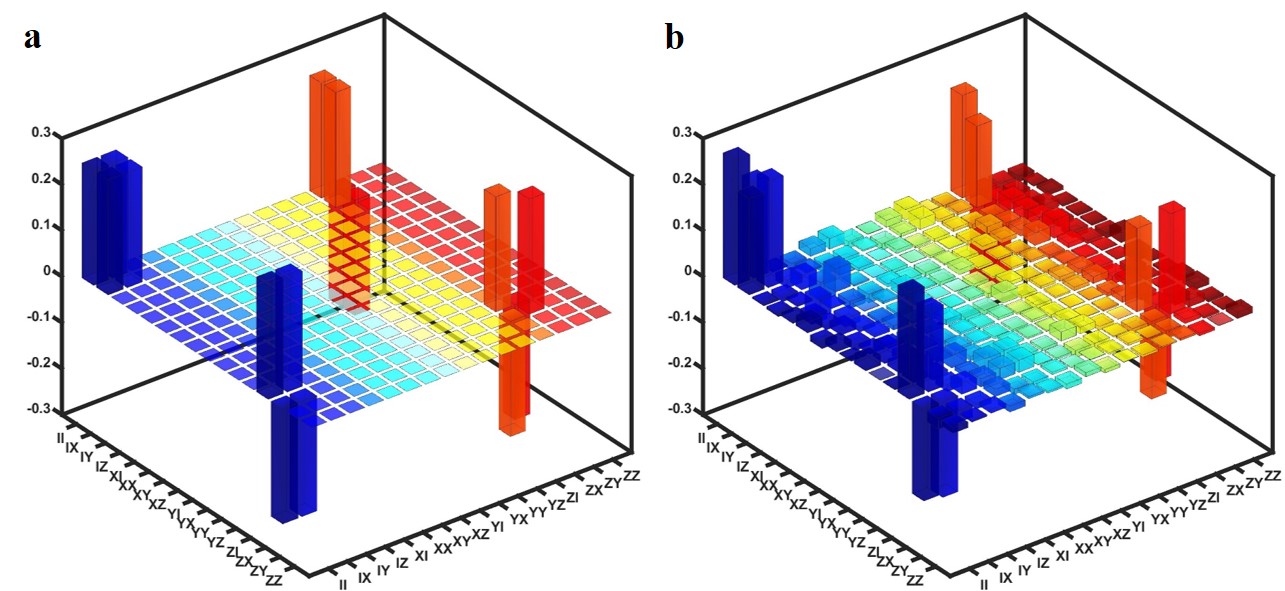}
\caption {Fig. 4.	Quantum process tomography measurement results. Real parts of the density matrices for (a) an ideal CNOT gate $\chi_{\rm{ideal}}$ and (b) the measured $\chi_{\rm{exp}}$. The imaginary part of $\chi_{\rm{exp}}$ is negligible. Here, $\bm{X}$, $\bm{Y}$, and $\rm{Z}$ represent the Pauli matrices $\sigma_{\rm{X}}$, $\sigma_{\rm{Y}}$, and $\sigma_{\rm{Z}}$, respectively.}
\label{fig4}
\end{figure*}

With a perfect CNOT gate, the input states are expected to be converted into the maximally entangled Bell states, that is,
\begin{equation}
\begin{aligned}
|\Phi^{\pm}\rangle=\frac{1}{\sqrt{2}}(\left|0\right\rangle_c\left|0\right\rangle_t\pm\left|1\right\rangle_c\left|1\right\rangle_t),\\
|\Psi^{\pm}\rangle=\frac{1}{\sqrt{2}}(\left|0\right\rangle_c\left|1\right\rangle_t\pm\left|1\right\rangle_c\left|0\right\rangle_t).
\end{aligned}
\end{equation}
The corresponding density matrix is denoted as $\hat\rho_{\rm{ideal}}=|\phi\rangle\langle\phi|$, where $|\phi\rangle=|\Phi^{\pm}\rangle,|\Psi^{\pm}\rangle$.
Fig. 3 shows the real parts of the reconstructed density matrices $\hat\rho_{\rm{mea}}$ and $\hat\rho_{\rm{ideal}}$ (crystal clear bars). The raw fidelity of the generated state, defined as $F_s=[\rm{Tr}(\sqrt{\sqrt{\hat\rho_{\rm{mea}}}\hat\rho_{\rm{ideal}}\sqrt{\hat\rho_{\rm{mea}}}})]^2$, were calculated for the different output states as $F_{|\Phi^{+}\rangle}=0.89\pm0.03$, $F_{|\Phi^{-}\rangle}=0.88\pm0.03$, $F_{|\Psi^{+}\rangle}=0.90\pm0.02$, $F_{|\Psi^{-}\rangle}=0.87\pm0.02$. The averaged fidelity was obtained as $\bar{F}_s=0.89\pm0.02$, which confirms the entangling function of the CNOT gate. The fidelity errors were obtained by a 100-times Monto Carlo calculation with the experimental data subjected to Gaussian statistics. In addition, the linear entropy and tangle were calculated to quantify the mixture and amount of entanglement for these states \cite{James2001,Wei2003,Coffman2000}. The details are provided in the Supplementary Materials (Section V).

The nonlocality nature of the output states was further verified using the Clauser-Horne-Shimony-Holt (CHSH) inequality \cite{Clauser1969}, \begin{equation}
\hat{S}(\hat{\rho})=E(a,b)-E(a,b')+E(a',b)+E(a',b').
\end{equation}
Here, $a$, $a'$ and $b$, $b'$ are two sets of measurement direction settings for different qubits. For a state that can be described by a local classical theory, the non-locality parameter $\hat{S}$ should be less than or equal to two. In other words, if the output state $\hat{\rho}$ cannot be described by a local classical theory, the inequality is violated. For entangled Bell states, quantum mechanics predicts a maximal value of $\hat{S}_{\rm{max}}=2\sqrt{2}$ with a selected measurement basis. In this study, the phases in the MZIs at the measuring part were individually set as $\{0^\circ,180^\circ,45^\circ,225^\circ\}$ for the control qubit and $\{22.5^\circ,202.5^\circ,67.5^\circ,247.5^\circ\}$ for the target qubit. In this way, $\hat{S}$ values for the different output states were obtained as $\hat{S}_{|\Phi^{+}\rangle}=2.54\pm0.05$, $\hat{S}_{|\Phi^{-}\rangle}=2.50\pm0.05$, $\hat{S}_{|\Psi^{+}\rangle}=2.51\pm0.05$, $\hat{S}_{|\Psi^{+}\rangle}=2.43\pm0.05$. The average value of the non-locality parameter is $\hat{S}_{\rm{Bell}}=2.50\pm0.05$, which violates the inequality by 10 standard deviations.


Finally, we performed quantum process tomography to fully characterize the quantum device.
We prepared 16 different input states from the set $\{\left|0\right\rangle,\left|1\right\rangle,\left|0\right\rangle+\left|1\right\rangle,\left|0\right\rangle+i\left|1\right\rangle\}^{\otimes 2}$and analyzed the output states. For each output state, 16 analysis bases were projected for the quantum state tomography measurement. Therefore, 256 projections were performed to fully characterize the two-qubit CNOT gate. The experimentally reconstructed process matrix $\chi_{\rm{exp}}$ is presented together with the ideal $\chi_{\rm{ideal}}$ in Fig. 4. Using the definition of the process fidelity $F=[\rm{Tr}(\sqrt{\sqrt{\chi_{\rm{exp}}}\chi_{\rm{ideal}}\sqrt{\chi_{\rm{exp}}}})]^2$ \cite{Crespi2011}, we obtained $F=0.82\pm0.01$, which is similar to or better than those reported in recent works \cite{Crespi2011,Politi2008,Shadbolt2012}.


The technological progress we have made here is multifaceted. First, as new multimode devices, the beam splitter and attenuator we designed and manufactured can also be used as basic components for various optical information processing applications. Second, the CNOT gate we achieved is the first and indispensable demonstration of a two-qubit logic gate with transverse mode-encoding. Together with the single qubit rotation operation realized previously \cite{Mohanty2017}, the CNOT gate makes universal quantum computation with transverse mode-encoded qubits possible. Compared to the chip size of path- and polarization-encoding CNOT gates based on dielectric waveguides (approximately 10$^5$-10$^7$ $\upmu$m$^2$) \cite{Crespi2011,Politi2008}, transverse-mode encoding enables us to further shrink the chip size to approximately 10$\times$160 $\upmu$m$^2$ and improve the chip stability when it becomes necessary to cascade multiple logic gates in large-scale circuits. The gate is promising for high-dimensional quantum information applications by introducing more transverse-modes. The gate is also compatible with the use of other degrees of freedom for encoding \cite{Feng2016} and offers a platform for hybrid encoding with diverse functions. Fully on-chip manipulation of transverse mode-encoded qubits is readily achievable with the integration of an entangled photon pair source using a multimode waveguide \cite{Feng2019}. It should be noted that the imperfect gate fidelity reported here is mainly due to the fabrication error, which resulted in insertion loss for the higher-order transverse mode. In principle, the higher-order transverse modes supported by a multimode waveguide can be transmitted with negligible loss through the use of expected advanced fabrication technologies.

In conclusion, we have demonstrated the manipulation of transverse mode-encoded qubits using a CNOT gate, which consists of newly developed multimode directional coupler and attenuator. The measured quantum state tomography shows that the entangled Bell states were generated by the gate operation with an average fidelity of $\bar{F}_s=0.89\pm0.02$, and that quantum locality was violated by 10 standard deviations. Quantum process tomography was also performed to fully characterize the device, which exhibited a process fidelity of $F_p=0.82\pm0.01$. The device is compact and extendable to higher dimensions, and paves the way for on-chip quantum information processing with multiple degrees of freedom and higher capacity.

This work was supported by the National Natural Science Foundation of China (NSFC) (Nos. 61590932, 11774333, 62061160487, 12004373, 91950205, 61961146003), National Science Fund for Distinguished Young Scholars (61725503), the Strategic Priority Research Program of the Chinese Academy of Sciences (No. XDB24030601), the National Key R \& D Program (No. 2016YFA0301700), Anhui Initiative in Quantum Information Technologies (No. AHY130300), Zhejiang Provincial Natural Science Foundation (LZ18F050001, LD19F050001), the Postdoctoral Science Foundation of China (No. 2020M671860) and the Fundamental Research Funds for the Central Universities. This work was partially carried out at the USTC Centre for Micro and Nanoscale Research and Fabrication.





\end{document}